\documentclass[prb,twocolumn]{revtex4-2}

\usepackage{graphicx}
\usepackage{amsmath}
\usepackage{amssymb}
\usepackage{bm}
\usepackage{bbm}
\usepackage{enumitem} 
\usepackage{float} 
\usepackage{empheq} 
\usepackage[title]{appendix}

\graphicspath{{./figures}}

\usepackage{xspace}
\newcommand{\wanBerri}{{\tt WannierBerri}\xspace}
\newcommand{\wBerri}{{\tt WB}\xspace}
\newcommand{\postwx}{{\tt postw90.x}\xspace}

\newcommand{\beq}{\begin{equation}}
\newcommand{\eeq}{\end{equation}}
\newcommand{\bea}{\begin{eqnarray}}
\newcommand{\eea}{\end{eqnarray}}
\newcommand{\beas}{\begin{subequations}\begin{eqnarray}}
\newcommand{\eeas}{\end{eqnarray}\end{subequations}}

\def\sumln{\sum_l\out\sum_n\inn}

\def\Bbar{\overline{B}}

\def\ket#1{\vert#1\rangle}
\def\bra#1{\langle#1\vert}
\def\ip#1#2{\langle#1\vert#2\rangle}
\def\me#1#2#3{\langle#1\vert#2\vert#3\rangle}

\def\wt#1{\widetilde{#1}}

\def\Re{{\rm Re\,}}
\def\Im{{\rm Im\,}}

\def\kp{\boldsymbol{\kappa}}
\def\kk{{\bf k}}
\def\qq{{\bf q}}
\def\dk{[d\kk]}
\def\KK{{\bf K}}
\def\rr{{\bf r}}
\def\RR{{\bf R}}

\def\inn{^{\text{occ}}}
\def\out{^{\text{unocc}}}

\def\ww{^{\text{W}}}

\def\abc{\epsilon_{\alpha\beta\gamma}}

\def\I{\cal I}
\def\T{\cal T}

\def\Hbar{\overline{C}^{\rm H}}

\def\Obar{\overline{\Omega}^{\rm H}}
\def\Abar{\overline{A}^{\rm H}}
\def\Bbar{\overline{B}^{\rm H}}

\makeatletter
\newcommand{\dwt}[1]{{%
  \mathpalette\double@widetilde{#1}%
}}
\newcommand{\double@widetilde}[2]{%
  \sbox\z@{$\m@th#1\widetilde{#2}$}%
  \ht\z@=.9\ht\z@
  \widetilde{\box\z@}%
}
\makeatother

\usepackage{xcolor}
\usepackage[colorlinks = true,
            linkcolor = blue,
            urlcolor  = blue,
            citecolor = blue,
            anchorcolor = blue]{hyperref}

\DeclareFixedFont{\ttb}{T1}{txtt}{bx}{n}{12} 
\DeclareFixedFont{\ttm}{T1}{txtt}{m}{n}{12}  

\usepackage{listings}
\usepackage{xcolor}

\definecolor{codegreen}{rgb}{0,0.6,0}
\definecolor{codegray}{rgb}{0.5,0.5,0.5}
\definecolor{codepurple}{rgb}{0.58,0,0.82}
\definecolor{backcolour}{rgb}{0.95,0.95,0.92}

\lstdefinestyle{mystyle}{
  backgroundcolor=\color{backcolour},  
  commentstyle=\color{codegreen},
  keywordstyle=\color{magenta},
  numberstyle=\tiny\color{codegray},
  stringstyle=\color{codepurple},
  basicstyle=\ttfamily\footnotesize,
  breakatwhitespace=false,         
  breaklines=true,                 
  captionpos=b,                    
  keepspaces=true,                 
  numbers=left,                    
  numbersep=5pt,                  
  showspaces=false,                
  showstringspaces=false,
  showtabs=false,                  
  tabsize=2
}

\begin{document}
    \title{ High performance Wannier interpolation of Berry curvature and related quantities: \wanBerri{} code}
\author{Stepan~S.~Tsirkin}
\email{stepan.tsirkin@uzh.ch}
\affiliation{Department of Physics, University of Zurich, Winterthurerstrasse 190, CH-8057 Zurich, Switzerland}

\begin{abstract}
Wannier interpolation is a powerful tool for performing Brillouin zone integrals over dense grids of $\kk$ points, which are essential to evaluate such quantities as the intrinsic anomalous Hall conductivity or Boltzmann transport coefficients. However, new physical problems and new materials create new numerical challenges, and computations with the existing codes become very expensive, which often prevents reaching the desired accuracy. In this article, I present a series of methods that  boost the speed of Wannier interpolation by several \textit{orders} of magnitude. 
They include a combination of fast and slow Fourier transforms, explicit use of symmetries and recursive adaptive grid refinement among others. The proposed methodology has been implemented in  the new python code \wanBerri, which also aims to serve as a convenient platform for the future development of  interpolation schemes for novel phenomena. 
\end{abstract}

\date{\today}

\maketitle

\section{Introduction}
Wannier functions \cite{Wannier-1937,marzari-review} (WFs) are a powerful tool for  evaluating various electronic properties of solids, ranging from electric polarization \cite{vanderbilt-prb-1993,kingsmith-1993,resta-RMP-1994} and orbital magnetization \cite{thonhauser-2005,ceresoli-2006,xiao-rmp10} to topological properties \cite{Soluyanov2011,Bradlyn2017-TQC,wu-WannierTools,Bouhon2019,Varnava2020}. 
Moreover, WFs provide a way to describe a group of energy bands in a crystal by a compact Hamiltonian, thus allowing a fast evaluation of the band structure at any  point of the Brillouin zone (BZ) without an extra call to the \textit{ab initio} code \cite{souza-disentangle}.
This procedure called Wannier interpolation is similar in spirit to the tight-binding method \cite{slater-koster}, but with significant advantage that it offers a systematic way of precise description of any number of bands without truncation of the hopping integrals. Moreover, not only the band energies, but also the wavefunctions and their derivatives over momentum space are precisely interpolated, while the tight-binding approach contains an unavoidable error due to the limited basis set.   
Wannier interpolation is particularly  useful  in the search for Weyl points in the band structure\cite{Gosalbez2015,wu-WannierTools},  and in the evaluation of momentum-space integrals  of rapidly varying functions. Such integrals appear, for example, in calculations of the anomalous Hall conductivity \cite{wang-prb06}, orbital magnetization \cite{lopez-prb12}, Boltzmann transport coefficients \cite{boltzwann}, and optical properties \cite{yates-2007,azpiroz-prb18}.
By now, Wannier interpolation schemes have been developed for many other properties that demand dense BZ sampling, such as electron-phonon coupling \cite{giustino-eph-wan,ponce-EPW-2016}, gyrotropic effects \cite{tsirkin-prb18}, and spin Hall conductivity \cite{qiao-SHC,ryoo-SHC}.

Due to their gauge dependence, WFs are strongly non-unique and may be constructed in multiple ways. The most popular technique is the maximal localization procedure \cite{marzari-prb97,marzari-review}, implemented in the well-established code Wannier90 \cite{MOSTOFI2008685,MOSTOFI20142309}, whose development is now driven by a broad community \cite{Pizzi_2020}. 
The construction of a good set of WFs requires some careful input from the user. This includes specifying a set of trial orbitals, that serve as an initial guess for the target WFs, and, if the bands of interest do not form an isolated group, choosing the  "disentanglement" energy windows \cite{souza-disentangle}. Recently there has been a significant progress in automated construction of WFs with minimal user intervention.  Different techniques have been proposed, such as the optimized projection functions method \cite{Mustafa2016}, the selected columns of the density matrix  method \cite{Damle-scdm,Damle-scdm2,Vitale2020}, and an automated method to choose trial orbitals  and energy windows \cite{zhang-highthrtopo,garrity2020database}. In addition, a database of Wannier Hamiltonians is currently being constructed \cite{JARVIS-WTB}.
These advances constitute significant steps towards employing Wannier interpolation for high-throughput automated calculations of electronic properties of solids.

Most of the Wannier interpolation schemes mentioned above have been implemented within the popular codes -- Wannier90 (namely its post-processing module \postwx) and WannierTools \cite{wu-WannierTools}. These codes, being well-established and widely adopted by the community, have quite broad functionalities.  However, new materials and new physical effects pose new numerical challenges, and calculation with those codes can become quite heavy. In particular, for a system with a large number of WFs and a complicated Fermi surface, it may be hard to achieve convergence with respect to momentum-space grid in a reasonable time. When it comes to high-throughput calculations, performance becomes even more important.

In this article, I present a series of methodological improvements that allow to improve dramatically the performance of Wannier interpolation method, without compromising its accuracy.  The proposed methodology is implemented in the new python code \wanBerri \cite{wberri-org} (\wBerri), which is freely available to install \cite{pypi-wberri} and open for contributors \cite{github-wberri}. The code has a broad functionality, and already a number of contributors. The present article covers only the core methodology of the code which procure its efficiency. 
More broad scope of the code can be found on its web page \cite{wberri-org} and will be detailed in future publications. Interesting to note that \wanBerri may be equally used for Wannier interpolation and tight-binding calculations, and also offers a convenient platform for development of new features. 
The name of the code is derived from Wannier functions and the Basque word ``berri'' which means ``new'' and enters local toponyms (e.g. Lekunberri, Ekainberri) 
\footnote{During the work on the code and on the manuscript, the author did not receive any funding from Basque institutions or foundations. Usage of Basque language is motivated by a coincidental similarity to the Berry phase, and by good memories of the years that the author spent in the Basque region.}. 

The high efficiency of the \wBerri code is achieved by the combination of  several methodological improvements. First, note that in a typical calculation using \postwx the bottleneck is the Fourier transform, which is implemented as a standard discrete Fourier transform. Therefore it looks appealing to use fast Fourier transforms (FFTs) \cite{cooley-tukey-1965-FFT,Heideman-1984-FFT} which are widely used in numerical calculations \cite{duhamel1990-FFT,vanLoan-book-FFT}. However, it is problematic to do so over a very dense grid of points, which may include up to $10^8$ points.  This issue is overcome in the present work by a mixed scheme employing both fast and ``slow'' Fourier transforms. Next,  the symmetries of the system may be used to reduce the evaluation only to the symmetry-irreducible points, and obtain the contributions from other points by applying symmetry operations. This also helps to render the result more symmetric (tensor components that should be equal or vanish will be exactly equal, or exactly vanish), even if the symmetries are slightly  broken due to numerical inaccuracies in wannierization.  Then  I introduce an adaptive refinement algorithm  that identifies points that give the largest contribution to the integral, and makes the grid more dense in the vicinity of such points. This helps to have a more accurate description near special points, where the integrand is rapidly changing -- for example near Weyl nodes or nodal lines. 
This is similar in spirit to the adaptive refinement scheme used in \cite{yao-ahc-Fe,wang-prb06}, but is more automatic and requires less input from the user. Finally,  I introduce  methods that drastically reduce the computational  cost of  the minimal-distance replica selection method (MDRS) \cite{Pizzi_2020}, as well as the cost  of scanning over multiple Fermi levels. 
 
 The article  is organized as follows. Section~\ref{sec:methods} describes the set of proposed improvements to the existing Wannier interpolation methodology.
 Section~\ref{sec:example} demonstrates the usage of the code for the 'textbook' example of  the anomalous Hall conductivity of bcc iron.
 Finally, in Sec.~\ref{sec:time} the performance of the \wBerri code is benchmarked against postw90 on the basis of that example. 
 Appendix~\ref{sec:capabilities} describes the list of functionalies implemented in the \wBerri code. Appendix~\ref{sec:matel} describes the routine to obtain the matrix elements that are not implemented in the interfaces of most \textit{ab initio} codes to Wannier90.

\section{Methods \label{sec:methods}}
    
    \subsection{General equations for Wannier interpolation\label{sec:wanfun}}
    The goal of this section is to introduce notation necessary for further discussion. For more details please refer to review \cite{marzari-review} and original articles cited therein. The problem of Wannier interpolation is stated in the following way. First we evaluate the energies $E_{n\qq}$ and wavefunctions $\psi_{n\qq}(\rr)\equiv e^{i\qq\cdot\rr}u_{n\qq}(\rr)$ from first principles on a rather coarse grid of $N_\qq=N_\qq^1\times N_\qq^2\times N_\qq^3$ wavevectors $\qq$ within the reciprocal unit cell.  Next we want to find the energies and wavefunctions at points on a denser grid of wavevectors $\kk$. Further we will consistently use $\qq$ and $\kk$ to denote the \textit{ab initio} and interpolation grids respectively.

For a group of entangled bands one can define a set of $J$ WFs defined as 
\begin{equation}
    \ket{\RR n}=\frac{1}{N_\qq}\sum_\qq  e^{-i\qq\cdot\RR} \sum_{m=1}^{{\cal J}_\qq} \ket{\psi_{m\qq}}V_{mn}(\qq),
\end{equation}
where ${\cal J}_\qq\ge J$ and $\RR$ are real-space lattice vectors. The matrices $V_{mn'}(\qq)$ contain all the information of the construction of WFs, and may be generated by Wannier90 code. They are constrained by $\sum_{m=1}^{{\cal J}_\qq} V^*_{mn}(\qq)V_{mn'}(\qq)=\delta_{nn'}$  and are chosen in such a way that the WFs are localized, which yields that the Bloch wavefunctions in the Wannier gauge 
\begin{equation}
    \ket{\psi_{n\kk}^{\rm W}} \equiv e^{i\kk\cdot\rr}\ket{u_{n\kk}^{\rm W}}\equiv  \sum_{\RR}e^{i\kk\cdot\RR}\ket{\RR n}  \label{eq:psiW}
\end{equation}
vary slowly with the $\kk$ vector, unlike the true wavefunctions. Now  let us see how WFs may be used to interpolate the band energies. First, one evaluates the matrix elements of the Hamiltonian
\begin{multline}
    H_{mn}(\RR)\equiv\frac{1}{N_\qq}\sum_\qq e^{-i\qq\cdot\RR} \me{\psi_{m\qq}^{\rm W}}{\hat{H}}{\psi_{n\qq}^{\rm W}}=\\
    =\frac{1}{N_\qq}\sum_\qq e^{-i\qq\cdot\RR}\sum_{l}V^*_{lm}(\qq)E_{l\qq}V_{ln}(\qq).
    \label{eq:fourier_q_to_R_H}
\end{multline}
Next, to obtain energies at an arbitrary point $\kk$ one needs to construct the Wannier Hamiltonian 
\begin{equation}
    H_{mn}^{\rm W}(\kk)=\sum_\RR H_{mn}(\RR)e^{i\kk\cdot\RR}, \label{eq:Hwann}
\end{equation}
which further may be diagonalized as 
\begin{equation}
   \sum_{mn} U_{ml}^*(\kk) H_{mn}^{\rm W}(\kk)U_{nl'}(\kk)=E_l(\kk) \delta_{ll'},
\end{equation}
where $U_{nl'}(\kk)$ are unitary matrices with columns corresponding to the eigenvectors of the Hamiltonian \eqref{eq:Hwann}. In a similar way, for any operator $\hat{X}$, for which the matrix elements are evaluated on the \textit{ab initio} grid, one  may obtain the real-space matrix elements
\begin{equation}
X_{mn}(\RR)\equiv\frac{1}{N_\qq}\sum_\qq e^{-i\qq\cdot\RR} X_{mn}\ww(\qq), \label{eq:fourier_q_to_R}
\end{equation}
where in a simple case (e.g. $\hat{X}=\boldsymbol{\sigma}$) 
\begin{equation}
X_{mn}\ww(\qq)= \sum_{ll'}V_{lm}^*(\qq) \me{\psi_{m\qq}}{\hat{X}}{\psi_{n\qq}} V_{l'n}(\qq), \label{eq:H_to_W}
\end{equation}
or if $\hat{X}$ involves momentum-space derivatives, (e.g. the position operator $\hat{r}_\alpha\equiv i\partial/\partial k\alpha$) may also involve matrix elements between neighbouring $\qq$ points (see \cite{wang-prb06,lopez-prb12} for details).
 Then the matrix elements may be interpolated to any $\kk$ point in the Wannier gauge by %
\begin{equation}
    X_{mn}^{\rm W}(\kk)=\sum_\RR X_{mn}(\RR)e^{i\kk\cdot\RR} , \label{eq:fourier_R_to_k}
\end{equation}
and further rotated to the Hamiltonian gauge
\begin{equation}
    \overline{X}_{mn}^{\rm H}(\kk)=\left( U^\dagger\cdot X^{\rm W}\cdot U \right)_{mn} . \label{eq:rotate_gauge}
\end{equation}
Note that equations \eqref{eq:fourier_q_to_R_H}, \eqref{eq:Hwann} are particular cases of \eqref{eq:fourier_q_to_R} and \eqref{eq:fourier_R_to_k}. Equation \eqref{eq:fourier_q_to_R} can be performed by means of FFT, and its result is periodic in $\RR$ with a  supercell formed by vectors $\mathbf{A}_i=\mathbf{a}_iN_\qq^i$, where $\mathbf{a}_i$ ($i=1,2,3$) are the primitive unit cell vectors. Among the equivalent $\RR$ vectors we  choose those belonging to the corresponding Wigner-Seitz (WS) supercell. If an $\RR$ vector belongs to the WS supercell boundary, we include all equivalent vectors on the boundary with the corresponding elements $X(\RR)$ divided by the degeneracy of the $\RR$ vector. Further, the  MDRS method (see Sec.~\ref{sec:replica}) may also slightly modify the set of $\RR$ vectors. 

As an example, the total Berry curvature of the occupied manifold is interpolated \cite{wang-prb06} via 
\begin{multline}
\Omega_\gamma (\kk) =   \Re\sum_n\inn \Obar_{nn,\gamma}
-2\abc \Re\sum_n\inn\sum_l\out D_{nl,\alpha}\Abar_{ln,\beta}  \\
 +\abc\Im\sum_n\inn \sum_l\out D_{nl,\alpha} D_{ln,\beta} ,
\label{eq:Berry-wanint}
\end{multline}
where the ingredients of the equation are obtained using eqs.~\eqref{eq:fourier_R_to_k}, \eqref{eq:rotate_gauge} starting from  $D_{nl,\alpha}\equiv\frac{\overline{H}_{nl,\alpha}^{\rm H}}{E_l-E_n}$, $H_\alpha^{\rm W}\equiv\partial_\alpha H^{\rm W}$, $A_{mn,\alpha}(\RR)\equiv\me{\bm{0}m}{\hat{r}_\alpha}{\RR n}$, $\overline{\Omega}_\gamma^{\rm W} \equiv\abc\partial_\alpha A^{\rm W}_\beta$, $\partial_\alpha\equiv \partial/\partial{k_\alpha}$. The anomalous Hall conductivity is evaluated as an integral
\begin{equation}
    \sigma_{\alpha\beta}^{\rm AHE}=-\frac{e^2}{\hbar}\abc\int \frac{d\kk}{(2\pi)^3}\Omega_\gamma(\kk).
    \label{eq:AHC}
\end{equation}

Note, that while the direct Fourier transform (\ref{eq:fourier_q_to_R}) is performed only once for the calculation, and is not repeated for the multiple $\kk$ points upon interpolation, the inverse Fourier transform (\ref{eq:fourier_R_to_k}) is repeated for every interpolation $\kk$ point. And in fact it presents the most time-consuming part of the calculation involving Wannier interpolation as implemented in the Wannier90 code.

    \subsection{Mixed Fourier transform \label{sec:FFT}}
    \begin{figure*}[th!]
    \centering
    \includegraphics[width=\textwidth]{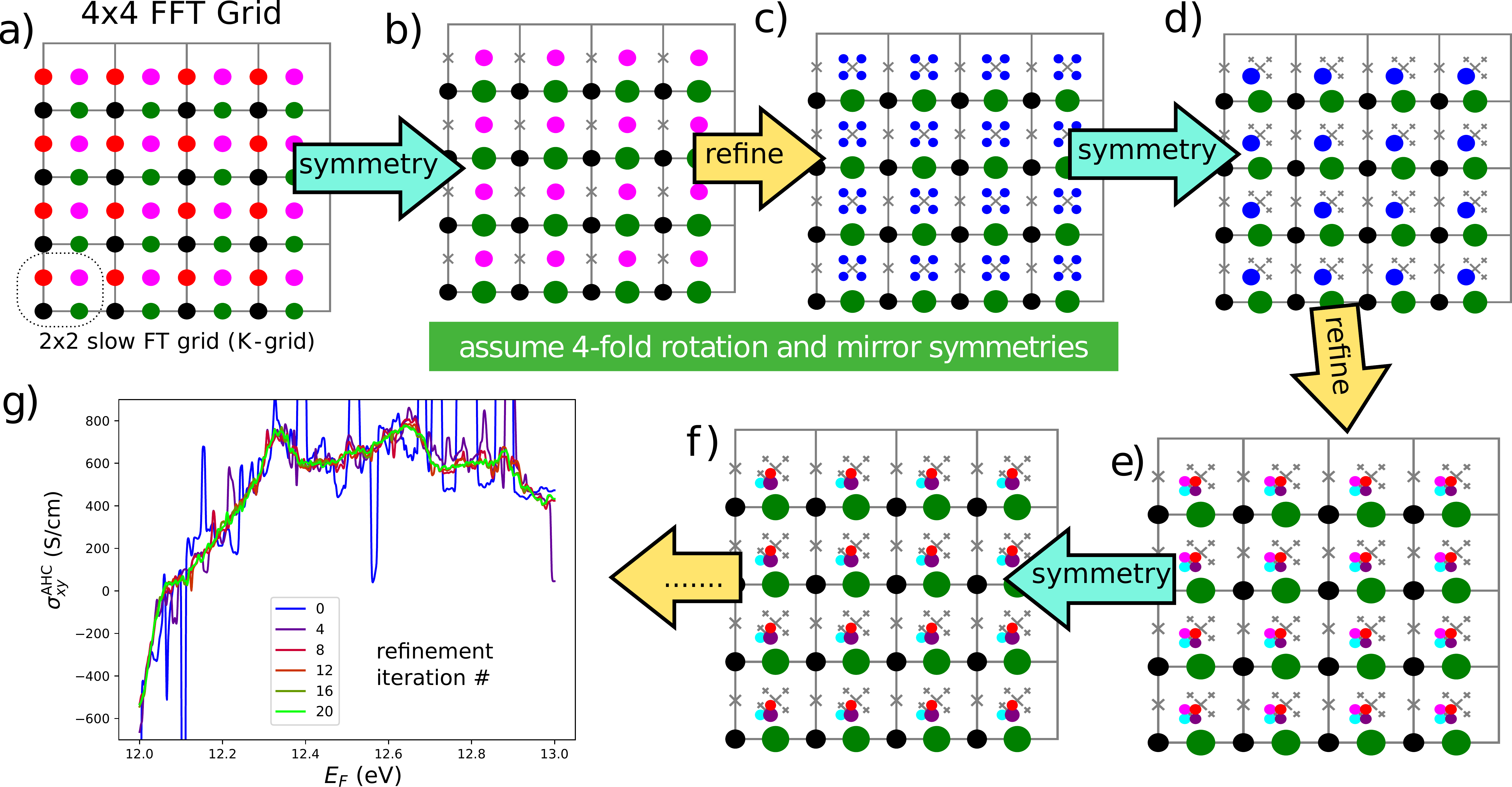}
    \caption{(a-f) Illustration of the procedure of mixed Fourier transform, adaptive refinement and use of symmetries. 2D picture is used for visualization purposes, while the code actually works in 3D.  The area of colored circles corresponds to the weight of the $\KK$-point, gray crosses denote the points with zero weight. See the text for detailed description. (g) AHC of bcc Fe, evaluated from a grid of 52$\times$52$\times$52 $\kk$-points and 20 recursive adaptive refinement iterations. }
    \label{fig:refinement}
\end{figure*}

In this section we will see how the evaluation of (\ref{eq:fourier_R_to_k}) may be accelerated. It is easy to see that 
the computation time of a straightforward discrete Fourier transform scales with the number of $\RR$ vectors and $\kk$ points as $t\propto N_\RR N_\kk$, and we are typically interested in a case $N_\kk \gg N_\RR$ ($N_\RR\approx N_\qq$).  

When the Fourier transform is done on a regular grid of $\kk$ points, it is usually appealing to use the FFT. 
For that one needs to place the $\RR$ vectors on a regular grid of size $N_\kk$, fill the missing spots with zeros and perform the standard FFT, which will scale as $t\propto N_\kk \log{N_\kk}$. However there are some dificulties with such FFT.
Mainly, because to perform FFT on a large grid implies storing the data for all $\kk$ points in memory at the same time, which becomes a severe computational limitation.  Also FFT does not allow to reduce computation to only the symmetry-irreducible $\kk$ points, and is more difficult to do in parallel. 
However there is a way to combine the  advantages of both the FFT and the usual discrete Fourier transform, leading to the concept of \textit{mixed Fourier transform}. 

We  want to evaluate (\ref{eq:fourier_R_to_k}) for a set of $\kk$ points.
\begin{equation}
\kk_{n_1,n_2,n_3}=\frac{n_1}{N_\kk^1}{\bf b}_1 +\frac{n_2}{N_\kk^2}{\bf b}_2 +\frac{n_3}{N_\kk^3}{\bf b}_3 ,   \label{eq:kgrid}
\end{equation}
where $0\le n_i< N_\kk^i$ -- integers ($i=1,2,3$), $N_\kk^i$ -- size of interpolation grid, ${\bf b}_i$ --- reciprocal lattice vectors. 
Now suppose we can factorize $N_\kk^i=N_{\rm FFT}^i N_\KK^i$ 
\footnote{This is always possible unless $N_\kk^i$ is a prime number. But for really dense grids, we can adjust $N_\kk^i$ a bit, to be factorizable in any way we want.} .
Then the set of points (\ref{eq:kgrid}) is equivalent to a set of points $\kk=\KK+\kp$, where 
\beas
\KK_{l_1,l_2,l_3}&=&\frac{l_1}{N_\kk^1}{\bf b}_1 +\frac{l_2}{N_\kk^2}{\bf b}_2 +\frac{l_3}{N_\kk^3}{\bf b}_3,  \label{eq:Kgrid}  \\
\kp_{m_1,m_2,m_3}&=&\frac{m_1}{N_{\rm FFT}^1}{\bf b}_1 +\frac{m_2}{N_{\rm FFT}^2}{\bf b}_2 +\frac{m_3}{N_{\rm FFT}^3}{\bf b}_3,    
\eeas
where $0\le l_i< N_\KK^i$, $N_\KK=\prod_i N_\KK^i$, $0\le m_i< N_{\rm FFT}^i$. 
This separation is illustrated in Fig~\ref{fig:refinement}(a), which shows a 2$\times$2 $\KK$-grid, each corresponding to 4$\times$4 FFT grid (dots of a certain color). 
Now for each $\KK$-point we can define 
\begin{equation}
    X_{mn}(\KK,\RR)\equiv X_{mn}(\RR)e^{i\KK\cdot\RR} \label{eq:XKR}
\end{equation}
and then \eqref{eq:fourier_R_to_k} reads as 
\begin{equation}
    X_{mn}^{\rm W}(\kk=\KK+\kp) = \sum_\RR X_{mn}(\KK,\RR)e^{i\kp\cdot\RR} \label{eq:XKk}
\end{equation}
The principle idea of mixed Fourier transform consists in performing the Fourier transform \eqref{eq:XKk} as FFT, while \eqref{eq:XKR} is performed directly. To perform the FFT we put all the $\RR$ vectors on a grid $N_{\rm FFT}^1\times N_{\rm FFT}^2\times N_{\rm FFT}^3$, and a vector $\RR=\sum_{i=1}^3 n_i\mathbf{a}_i$ is placed on a slot with coordinates $\wt{n}_i= n_i\,{\rm mod}\,N_{\rm FFT}^i$ ($n_i$ are both positive and negative integers, while $0\le \wt{n}_i<N_{\rm FFT}^i$). It is important to choose the FFT grid big enough, so that two different $\RR$ vectors are not placed on the same slot in the grid.

The advantages of this approach are the following. 
First, the computational time scales as $t_1\propto N_\KK N_\RR $ for \eqref{eq:XKR} and $t_2\propto N_\KK N_{\rm FFT}\log N_{\rm FFT} $ for \eqref{eq:XKk}. Because it is required that $N_{\rm FFT}\ge N_\RR$ (to fit all $\RR$-vectors in the FFT box),
we have $t_1 \le t_2 \propto N_\kk\log N_{\rm FFT}$ (in practice it occurs that $t_1 \ll t_2$), 
which scales better then both the Fast and 'slow' Fourier transforms. Next, we can perform Eqs.~\eqref{eq:XKR} and \eqref{eq:XKk} independently for different $\KK$-points. 
This saves us memory, and also offers a simple parallelization scheme. 
Also we can further restrict evaluation only to symmetry irreducible $\KK$-points (Sec.~\ref{sec:symmetry}) and also perform adaptive refinement over $\KK$-points (Sec.~\ref{sec:adaptive}). 

Moreover, the evaluation time of a mixed Fourier transform only logarithmically depends on the size of the \textit{ab initio} grid (recall that $N_{\rm FFT}\sim N_\RR\sim N_\qq$), while for the slow Fourier transform, the dependence is linear. However, in practice we will see (sec.~\ref{sec:time}) that the Fourier transform in the present implementation consumes only a small portion of computational time, and therefore the overall computational time is practically independent of the size of the \textit{ab initio} grid.

   \subsection{Symmetries \label{sec:symmetry}}
    When we integrate some quantity over the BZ, at every  $\KK$-point (after summing over $\kp$ points) we obtain the result as a rank-$m$ tensor  $X_{i_1,\ldots,i_m}(\KK)$, for example the berry curvature vector $\Omega_\gamma$ or the conductivity tensor $\sigma_{xy}$.  Then the  BZ  integral is expressed as a sum
\begin{equation}
    {\cal X}=\sum_\KK^{\rm all}  X(\KK)w_\KK \label{eq:sumK}
\end{equation}
and we initially set $\{\KK\}$ as a regular grid \eqref{eq:Kgrid}  and $w_\KK=1/N_\KK$. Suppose $G$ is the magnetic point group of the system \footnote{Because $X(\KK)$ is invariant under translations, here we are interested in the point group, rather then space group.}. We define the set of symmetry-irreducible $\KK$ points $\rm irr$ as a a set of points that $\forall \KK,\KK'\in{\rm irr}$,  $\forall g\in G$ holds $g\KK\neq\KK'$, unless $g=E$ (identity). Then we can rewrite the sum \eqref{eq:sumK} as 
\begin{equation}
      {\cal X}=\sum_\KK^{\rm all}  g_\KK X(g_\KK^{-1}\KK)w_\KK
      \label{eq:sumK-split} 
\end{equation}
where we choose $g_\KK$ such that $g_\KK^{-1}\KK\in{\rm irr}$ (this choice may be not unique), and obviously $g_\KK=E$ for $\KK\in{\rm irr}$. Thus, only the irreducible $\KK$ points need to be evaluated.  Next, to make sure that the result respects the symmetries, despite possible numerical inaccuracies, we symmetrize the result as:
\beq
    {\cal\widetilde X} = \frac{1}{|G|}\sum_f^{G} f {\cal X}.   \label{eq:symmetrize}  
\eeq
Note, that ${\cal\widetilde X}={\cal X}$ if the model respects the symmetry precisely (e.g. when symmetry-adapted WFs \cite{sakuma-SAWF} are used).
Combining \eqref{eq:sumK-split} and \eqref{eq:symmetrize} and using $\sum_f^{G} f\cdot g_\KK = \sum_f^{G} f$ we get 
\beq
    {\cal\widetilde X}= \frac{1}{|G|}\sum_f^{G} f \left[\sum_\KK^{\rm irr}  X(\KK) \left( \sum_{\KK'}^{G\cdot\KK} w_{\KK'} \right) \right] , \label{eq:symmetrize-final}  
\eeq
where $G\cdot\KK$ denotes the orbit of $\KK$ under action of group $G$. The latter equation reflects the implementation in the \wBerri code. Starting from a regular grid of $\KK$ points we search for pairs of symmetry-equivalent points. Whenever such a pair is found, one of the points is excluded and it's weight is transferred to the other point. Compare Figs.~\ref{fig:refinement}(a) and (b): the red points are removed and their weight is moved to green points. Thus we end with a set of irreducible $\KK$-point with weights $\wt w_\KK=\sum_{\KK'}^{G\cdot\KK} w_{\KK'}$. Next we evaluate $X(\KK)$ (employing the corresponding interpolation scheme) only at symmetry-irreducible $\KK$-points. Note, that although some $\kk$-points corresponding to the same $\KK$-point (same color in Fig.~\ref{fig:refinement}) are equivalent, we have to evaluate them all to be able to use the FFT. Finally, after summation, we symmetrize the result. 
The described procedure achieves two goals: (i) reduce the computational costs and (ii) make the result precisely symmetric, even if the WFs are not perfectly symmetric. In the present example we managed to obtain highly symmetric WFs (although without employment of symmetry-adapted WFs method), and therefore the symmetrization procedure does not change the result (within relative accuracy $\sim 10^{-5}$). However, for complex materials such quality of WFs is not always easy to achieve.

    \subsection{Recursive adaptive refinement\label{sec:adaptive}}

It is well known that in calculations of quantities involving Berry curvature or orbital moments, one performs integration over $\kk$-space of a function that rapidly changes with $\kk$. 
As a result, small areas of $\kk$-space give the major contribution to the integral. 
Such areas often appear in the vicinity of Weyl points, nodal lines, as well as avoided crossings. 
To accelerate convergence with respect to the number of  $\kk$ points, we utilize adaptive mesh refinement similar to Refs. \cite{yao-ahc-Fe,wang-prb06}. 
The authors of \cite{yao-ahc-Fe,wang-prb06} assumed a pre-defined threshold, and the $\kk$-points yielding Berry curvature above the threshold were refined. This is inconvenient because one needs a good intuition to guess an optimal value for this threshold, because it depends both on the quantity one wants to calculate, and the material considered.

In \wBerri it is implemented in a way that does not require initial guess from the user. 
This procedure, in combination with symmetrization described above, is illustrated in Fig.~\ref{fig:refinement} in two dimensions (2D), while the actual work in 3D is described below.  
After excluding symmetry-equivalent $\KK$-points (Fig.~\ref{fig:refinement}b) the results are evaluated for every $\KK$ point and stored. We assume that initially each $\KK$ point has weight $\wt w_\KK$ and corresponds to a volume defined by vectors $\mathbf{c}_\KK^i=\mathbf{b}_i/N_\kk^i$ centered at $\KK$. Then we pick a few "most important $\KK$-points". 
The criteria of importance may be different - either the Maximal value for any $E_F$, or maximal value summed over all $E_F$, or yielding most variation over the $E_F$ (if the evaluated quantity is a function of Fermi level $E_F$). 
Suppose we selected the magenta point.  
Then those points are refined --- replaced with 8 points around it with coordinates
\begin{equation}
    \KK'=\KK\pm\frac{\mathbf{c}_\KK^1}{4}\pm\frac{\mathbf{c}_\KK^2}{4}\pm\frac{\mathbf{c}_\KK^3}{4}, 
\end{equation}
where all combinationgs of $\pm$ signs are used. In Fig.~\ref{fig:refinement}c 4 new blue $\KK$-points in the 2D case.  
The weight and volume of the initial point is distributed over the new points, thus $w_{\KK'}=\wt w_\KK/8$ and $\mathbf{c}_{\KK'}^i=\mathbf{c}_{\KK}^i/2$. 
Then the symmetrization is applied again (the four blue points are connected by 4-fold rotation)  to exclude the equivalent points, and the weight of the equivalent points is collected on the remaining point, while the vectors $\mathbf{c}_{\KK'}^i$ are not changed. After the new $\KK$-points are evaluated, we go to the next iteration of refinement. 
On each iteration any point may be refined, including both those from the initial regular grid, and those created during previous refinement iterations. 
The procedure stops after the pre-defined number of iterations was performed. 
Figure~\ref{fig:refinement}(g) shows how undesired artificial peaks  of the the AHC curve are removed iteration by iteration, yielding a smooth curve (See sec.~\ref{sec:example} for details). 

    \subsection{ Minimal-distance replica selection method \label{sec:replica}}
    The MDRS method \cite{Pizzi_2020} allows to obtain a more accurate Wannier interpolation, in particular when moderate $\qq$-grids are used in the \textit{ab initio} calculations. With MDRS method the Fourier transform \eqref{eq:fourier_R_to_k} is modified in the following way:
\begin{equation}
    X_{mn}^{\rm W}(\kk)=\sum_\RR\frac{1}{{\cal N}_{mn\RR}} X_{mn}(\RR)\sum_{j=1}^{{\cal N}_{mn\RR}} e^{i\kk\cdot\left(\RR+\mathbf{T}_{mn\RR}^{(j)}\right)} ,\label{eq:replica}
\end{equation}
where $\mathbf{T}_{mn\RR}^{(j)}$ are ${\cal N}_{mn\RR}$ lattice vectors that minimise the distance $|\rr_m-(\rr_n+\RR+{\bf T})|$ for a given set $m,n,\RR$. However, the evaluation of \eqref{eq:replica} is quite slower than \eqref{eq:fourier_R_to_k}, because every $\kk,m,n,\RR$ an extra loop over $j$ is needed. Therefore calculations employing MDRS  in {\tt postw90.x} (which is enabled by default) takes more time.  Instead it is convenient to re-define the modified real-space matrix elements as 
\begin{equation}
    \wt X_{mn}(\RR) = \sum_{\RR'} \frac{1}{{\cal N}_{mn\RR'}} X_{mn}(\RR') \sum_{j=1}^{{\cal N}_{mn\RR'}}   \delta_{\RR,\RR'+\mathbf{T}_{mn\RR'}^{(j)}}\label{eq:replica1}
\end{equation}
only once for the calculation, and then  the transformation to $\kk$-space is performed via  
\begin{equation}
    X_{mn}^{\rm W}(\kk)=\sum_\RR e^{i\kk\RR} \wt X_{mn}(\RR). \label{eq:replica2}
\end{equation}
Note, that the set of $\RR$ vectors in \eqref{eq:replica1} is increased compared to the initial set of vectors in \eqref{eq:fourier_q_to_R} in order to fit all nonzero elements $\wt X_{mn}(\RR)$
Equation \eqref{eq:replica2} having essentially same form as \eqref{eq:fourier_R_to_k},  can be evaluated via mixed Fourier transform, as described in Sec.~\ref{sec:FFT}.

Thus the MDRS method implemented in \wBerri via Eqs.~\eqref{eq:replica1}-\eqref{eq:replica2}, and  has practically no extra computational cost, while giving notable accuracy improvement. 

    \subsection{Scanning multiple Fermi levels \label{sec:fermisea}}
    It is often needed to study anomalous Hall conductivity (AHC) not only for the pristine Fermi level $E_F$, but considering it  as a free parameter $\epsilon$. 
On the one hand it gives an estimate of the accuracy of the calculation, e.g. sharp spikes may indicate that the result is not converged. On the other hand $\epsilon$-dependence gives access to the question of the influence of doping and temperature, and also allows calculation of anomalous Nernst effect \eqref{eq:ANE}. As implemented in \postwx, evaluation of multiple Fermi levels has a large computational cost.
However there is a way to perform the computation of AHC  for multiple Fermi levels without extra computational costs. To show this let's rewrite \eqref{eq:Berry-wanint}, \eqref{eq:AHC} as $\sigma_{\alpha\beta}(\epsilon)=-\abc\frac{e^2}{\hbar}\Omega_\gamma(\epsilon)$, where $\Omega_\gamma(\epsilon)=\sum_\KK w_\KK \Omega_\gamma(\KK,\epsilon)$ and
\beq
\Omega (\KK,\epsilon) = \sum_{\kp}\left( \sum_n^{O(\kk,\epsilon)} P_n(\kk) + \sum_l^{U(\kk,\epsilon)}\sum_n^{O(\kk,\epsilon)} Q_{ln}(\kk) \right),
\label{eq:Osum-o-uo}
\eeq
where $\kk=\KK+\kp$, the definitions of $P_n$ and $Q_{ln}$ straightly follow from \eqref{eq:Berry-wanint}, and we omit the cartesian index $\gamma$ further in this subsection.
Now suppose we want to evaluate $\Omega(\epsilon_i)$ for a series of Fermi levels $\epsilon_i$.
For different $\kk$-points and Fermi levels $\epsilon$  the sets of occupied $O(\kk,\epsilon)$ and unoccupied states $U(\kk,\epsilon)$ change and repeating this summations many times may be computationally heavy. 
Instead we note that when going from one Fermi level $\epsilon_i$ to another $\epsilon_{i+1}$ only a few states at a few $\kp$-points change from unoccupied to occupied. Let's denote the set of such $\kp$-points as $\delta \kappa_i$
then, the change of the total Berry curvature is 
\begin{widetext}
\begin{multline}
\delta{\Omega}_i \equiv  \Omega(\epsilon_{i+1})-{\Omega}(\epsilon_i)=\\=
 \sum_\kk^{\delta \kappa_i} \left(  \sum_n^{O(\kk,\epsilon_{i+1})} P_n(\kk) + \sum_l^{U(\kk,\epsilon_{i+1})}\sum_n^{O(\kk,\epsilon_{i+1})} Q_{ln}(\kk) -
  \sum_n^{O(\kk,\epsilon_{i})} P_n(\kk) - \sum_l^{U(\kk,\epsilon_{i})}\sum_n^{O(\kk,\epsilon_{i})} Q_{ln}(\kk) \right)= \\=
\sum_\kk^{\delta \kappa_i} \left( \sum_n^{\delta O_i(\kk)} P_n + 
\sum_l^{U(\kk,\epsilon_{i+1})}\sum_n^{\delta O_i(\kk)} Q_{ln}(\kk)
-\sum_l^{\delta O_i(\kk)}\sum_n^{O(\kk,\epsilon_i)} Q_{ln}(\kk) \right),
\label{eq:deltamu}
\end{multline}
\end{widetext}
where $\delta O_i(\kk)\equiv O(\kk,\epsilon_{i+1})-O(\kk,\epsilon_{i})$. Note that if the step $\epsilon_{i+1}-\epsilon_i$ is small, then $\delta \kappa_i$ and $\delta O_i(\kk)$ include only few elements, if not empty. Hence the evaluation of \eqref{eq:deltamu} will be very fast. Thus, the full summation \eqref{eq:Osum-o-uo} is needed only for the  first Fermi level.  

In a similar way this approach may be  applied to orbital magnetization and other Fermi-sea properties. E.g. the orbital magnetization may be written as 
\bea
M_\gamma (\kk) &=& \sum_n\inn\Re\left[\Hbar_{nn,\gamma} + E_n\Obar_{nn,\gamma}  \right] - \nonumber \\
&&-2\abc\sumln\Re\left[D_{nl,\alpha}(\Bbar_{ln,\beta}+ \Abar_{ln,\beta}E_n)\right] \nonumber\\
&&+\abc\Im\sumln D_{nl,\alpha} (E_l+E_n) D_{ln,\beta}
\label{eq:Morb-wanint}
\eea
where $C_{mn,\gamma}(\RR)\equiv\abc\me{\bm{0}m}{r_\alpha\cdot\hat{H}\cdot(r_\beta-R_\beta)}{\RR n}$, $B_{mn,\beta}(\RR)\equiv\me{\bm{0}m}{\hat{H}\cdot(r_\beta-R_\beta)}{\RR n}$  and the other ingredients were explained under \eqref{eq:Berry-wanint}.
Equation~\eqref{eq:Morb-wanint} is written following the approach of Ref.~\onlinecite{lopez-prb12}, but the result has a different form, which can be  straightforwardly processed by analogy with \eqref{eq:Osum-o-uo} and  \eqref{eq:deltamu}, where  the first line of \eqref{eq:Morb-wanint} expresses $P_n(\kk)$ while the second and third lines correspond to $Q_{ln}(\kk)$.

\section{Example: AHC of bcc iron \label{sec:example} }
In this section the usage of the \wanBerri code is demonstrated on a simple example --  anomalous Hall conductivity of bcc iron.
First we performed the ab-initio calculations using the QuantumEspresso (QE) code \cite{QE-2020} on a grid 8$\times$8$\times$8 $\qq$-points, fixing the magnetization along [001] axis. Next, we construct 18 WFs describing the conduction band of Fe. These two steps follow exactly the Tutorial\#18 of Wannier90 (W90) \cite{wannier.org}, and the reader is addressed to the documentation of Wannier90 and QE for details.  

After that, the calculation is performed by the following short python script. First, we import the needed packages:
\begin{lstlisting}[language=Python,name={tutorial}]
import wannierberri as WB
import numpy as np
\end{lstlisting}
Then we read the information about the system and  WFs:
\begin{lstlisting}[language=Python,name={tutorial}]
system=WB.System_w90('Fe',berry=True)
\end{lstlisting}
from  files {\tt Fe.chk}, {\tt Fe.eig}, {\tt Fe.mmn} 
\footnote{the first is written by Wannier90, the other two by the interface of the ab initio code (e.g. pw2wannier90.x)}, or we can read all information from a file 
 {\tt Fe\_tb.dat}, which is also written by Wannier90, or maybe composed by user from any tight-binding model:
\begin{lstlisting}[language=Python,numbers=none]
system=WB.System_tb('Fe_tb.dat',getAA=True)
\end{lstlisting}
Next, we define the symmetries that we wish to take into account.
In the \textit{ab initio} calculation we have specified the magnetization along the $z$ axis, hence the symmetries that are preserved are inversion $\I$, 4-fold rotation around the $z$ axis $C_{4z}$, and a combination of time-reversal $\T$ and 2-fold rotation around the $x$ axis $C_{2x}$.
Here we need only the generators of the symmetry group. 
\begin{lstlisting}[language=Python,name={tutorial}]
system.set_symmetries([
    'Inversion','C4z','TimeReversal*C2x'])
\end{lstlisting}
The other symmetries will be automatically obtained by taking products of these generators, for example, the mirror is $M_z=(C_{4z})^2\cdot \I$. 

Next we need to set the grids of $\kk$, $\KK$ and $\kp$ points. Most conveniently it can be done by setting the {\tt 'length'} parameter (in \AA):
\begin{lstlisting}[language=Python,name={tutorial}]
grid=WB.Grid(system,length=100)
\end{lstlisting}
This will guarantee the grid to be consistent with the symmetries, and the spacing of $\kk$-points will be  $\Delta k\approx \frac{2\pi}{\rm length}$. In this particular case, the reciprocal lattice vectors have length $|\mathbf{b}_i|=3.1$ \AA$^{-1}$, hence the suggested grid size is $\frac{{\rm length}\cdot|\mathbf{b}_i|}{2\pi}\approx49$. However, a minimal FFT grid $13\times13\times13$ is needed to fit all $\RR$ vectors. Therefore the grid size are adjusted to  $52\times52\times52$ points ($13\times13\times13$ $\kp$-grid, $4\times4\times4$ $\KK$-grid).

Next, we want to integrate the Berry curvature to get the AHC. This is done by the \lstinline{WB.integrate} method.  
\begin{lstlisting}[language=Python,name={tutorial}]
WB.integrate(system, grid, 
            Efermi=np.linspace(12.,13.,1001), 
            smearEf=10, # 10K
            quantities=["ahc","cumdos"],
            numproc=16,
            adpt_num_iter=20,
            fout_name="Fe")
\end{lstlisting}
 and in addition to AHC we evaluate the cumulative density of states (cDOS) \eqref{eq:cDOS}. We consider Fermi level as a free parameter, scanning  over a set of Fermi levels from 12 to 13 eV with a step of 1 meV, and a small smearing over the Fermi level corresponding to temperature 10K ($\sim1$ meV) is used.  
 It is known, that in the BZ integration, some $\kk$ points may give a large contribution to the integral. This is especially strong for Berry curvature, which blows up near band degeneracies and avoided crossings, that fall close to the Fermi level. This is reflected as huge spikes in the $E_F$-resolved curves -- see blue curve in Fig~\ref{fig:refinement}(g). 
 To make the calculation more precise around such points, an adaptive recursive refinement algorithm is used, and we set the number of iterations to 20.  
 The integration is done in parallel over $\KK$-points by means of the \lstinline{multiprocessing} \cite{multiprocessing} module and the parameter {\tt 'numproc'} specifies that a \lstinline{Pool} of 16 worker processes is used.

From the cDOS  we can find the precise position of the Fermi level  $E_F=12.610$ eV --- the energy at which the cumulative DOS reaches 8 electrons per unit cell. This is more accurate, then the result evaluated from a coarse \textit{ab initio} grid. Next, it is instructive to plot the AHC after each iteration. In Fig.~\ref{fig:refinement}g one can see that already after 20 iterations  the chaotic peaks are removed, and we can get a reasonably sooth curve, although we have started from a rather coarse grid of only 52$\times$52$\times$52 $\kk$-points. 

This short example demonstrates that the calculations with \wBerri may be run with a few lines of Python script. Appendix~\ref{sec:capabilities} describes more of the implemented functionality, and more options are under development or under testing. For more detailed and updated information please refer to the online documentation at \cite{wberri-org}. 

\section{Computation time \label{sec:time}}
In this section we will compare the time for the calculations of anomalous Hall conductivity using \postwx and \wanBerri. We will take the example of bcc Fe and vary different parameters. Calculations were performed on identical 32-core virtual nodes of the ScienceCloud cluster at University of Z\"urich. The nodes are based on AMD EPYC 7702 64-Core Processors with frequency 2GHz and 128 GB RAM per node, and one node was used per task.

The computation consists of two phases. First, some preliminary operations are done.
Those include reading the input files and performing Fourier transform from \textit{ab initio} grid $\qq$ to real-space vectors $\RR$  : eqs.~\eqref{eq:fourier_q_to_R_H}, \eqref{eq:fourier_q_to_R} and \eqref{eq:H_to_W}.
This operation takes in \wBerri (\postwx) between 2 (3) seconds for the small 
$\qq$ grid $4\times4\times4$ and 2 (3) minutes for a large grid of $16\times16\times16$. 
This time is mostly taken by reading the large formatted text file {\tt Fe.mmn}, and it is done only once and does not scale with the density of the interpolation grid.  
In \wBerri this is done in the constructor of the {\lstinline{System_w90}} class, and the object can be saved on disk using a \lstinline{pickle} module, so that this operation does not repeat for further calculations. 

Next comes the interpolation part itself, for which the evaluation time  scales linearly with the number of $\kk$-points used. Further the time for an interpolation grid 200$\times$200$\times$200 is given, which is a rather good grid to make an accurate calculation for this material. 

\begin{figure}
    \centering
    \includegraphics[width=\columnwidth]{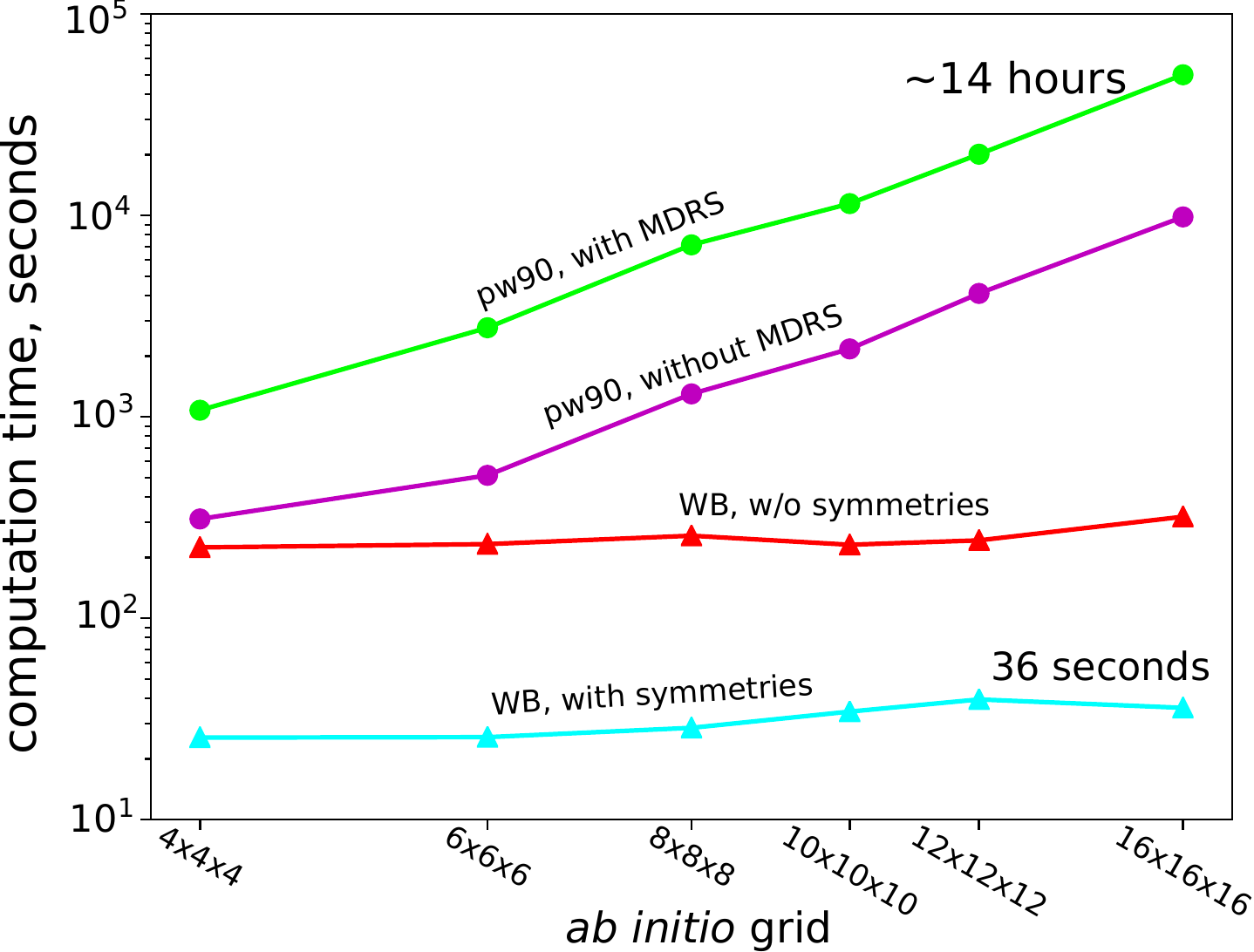}
    \caption{Computational time for AHC using \wBerri (triangles) and \postwx (circles) for different \textit{ab initio} grids. For \postwx the calculations are done with (green) and without (purple) MDRS. For \wBerri the calculations are done with (cyan) and without (red) use of symmetries.}
    \label{fig:timing}
\end{figure}
We start with comparing time  with the MDRS switched off and without use of symmetries in \wBerri. As can be seen in Fig.~\ref{fig:timing}, for a small  $\qq$-grid 4$\times$4$\times$4 \wBerri is just slightly faster then \postwx. However, for dense $\qq$-grids the computational time of \postwx grows linearly with the number of $\qq$ points, while in \wBerri it stays almost the same. 
This happens because in \postwx the Fourier transform is major time-consuming routine. On the other hand, in \wBerri, although cost of the mixed Fourier transform is expected to grow logarithmically with the \textit{ab initio} grid, we do not see it because Fourier transform amounts only to $\sim 10$\% of the computational time. 

Next, we switch on the MDRS method, and the computational time of \postwx grows by a factor of 5. On the other hand the computational time of \wanBerri does not change (not shown).

Finally let's switch on the use of symmetries in $\wBerri$. Thus the computational time decreases by a factor of 8. In the ultra-dense grid limit one would expect the speedup to be approximately equal to the number of elements in the group -- 16 in the present example, due to exclusion of symmetry-equivalent $\KK$-points. But this does not happen, because we use an FFT grid of 25$\times$25$\times$25 $\kp$-points, hence the $\KK$-grid is only $8\times8\times8$, and a considerable part of $\KK$-points are at high-symmetry positions. Therefore they do not have symmetric partners to be excluded from the calculation. 

\begin{figure}
    \centering
    \includegraphics[width=\columnwidth]{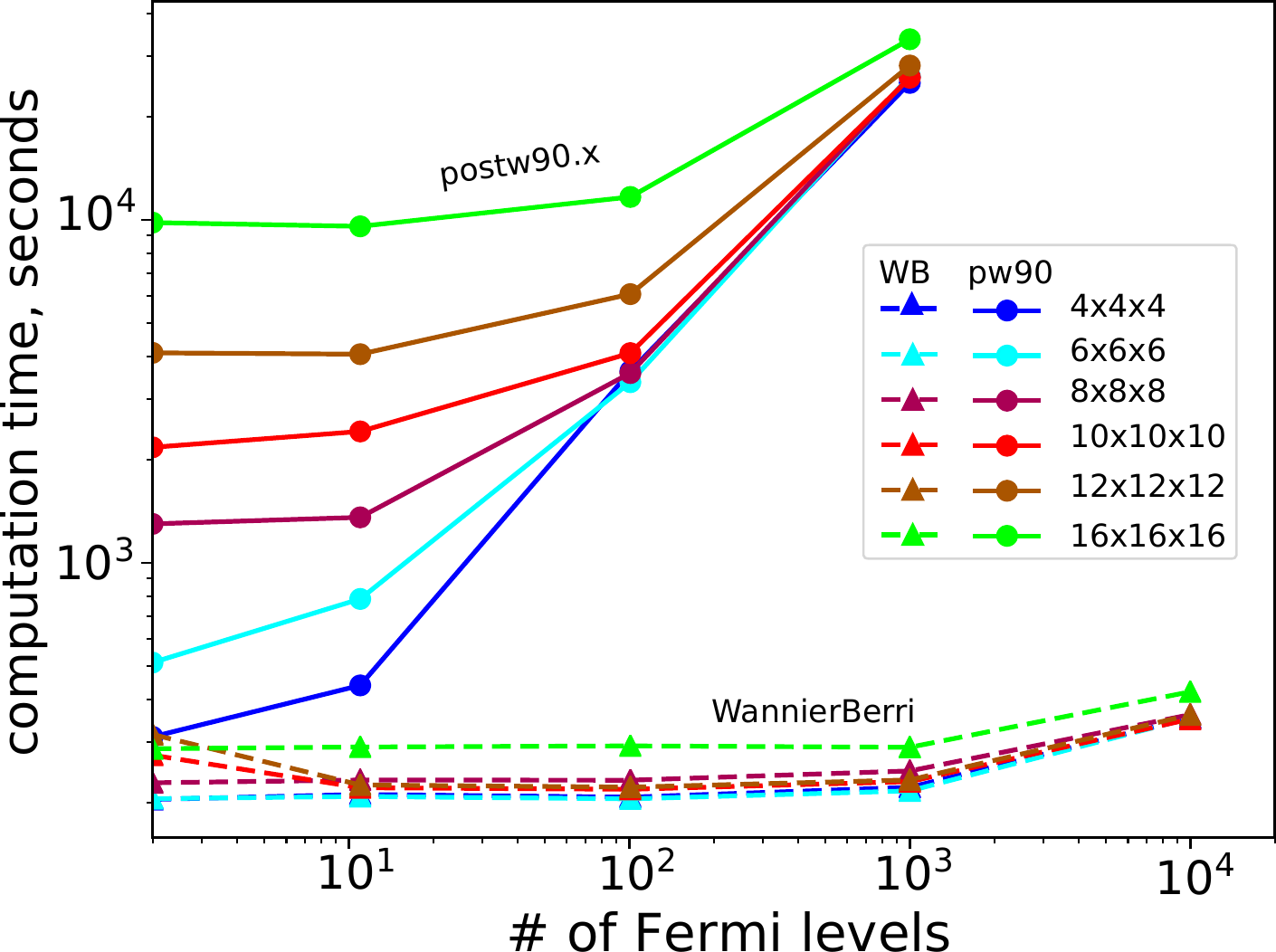}
    \caption{Computational time for scanning multiple Fermi levels using \wBerri (dashed lines) and \postwx (pw90, solid lines) for different \textit{ab initio} grids. MDRS method and symmetries are disabled here. }
    \label{fig:timing-fscan}
\end{figure}

Thus we can see that the difference in computational time with \postwx and \wBerri reaches 3 orders of magnitude for this example. Note that the examples above were performed only for the pristine Fermi level. Now let's see what happens upon scanning the Fermi levels (Fig.~\ref{fig:timing-fscan}). In \wBerri the computational time remains practically unchanged when we use upto $N_\epsilon\approx1000$ Fermi levels, and only start to grow considerably at $N_\epsilon\sim 10^4$. On the other hand in \postwx the computational time significantly grows with $N_\epsilon$, which is especially remarkable for small $\qq$ grids, where the growth becomes linear already from $N_\epsilon\sim10$. For denser $\qq$ grids the fixed amount of time (independent of $N_\epsilon$) is larger, so the linear growth starts at higher $N_\epsilon$.

In this section we did not use the adaptive refinement procedure. However when one starts from a rather large grid of $\KK$-points, the new $\KK$-points coming from the refinement procedure constitute only a small portion of the initial grid, and hence do not contribute much into computation time.  

In this section we have shown that the methods suggested in this article help to significantly reduce the computation time from days to minutes.  However, bcc iron is a simple example with only 1 atom per unit cell, and only 18 WFs are needed.  More complicated systems  will require more time. 
For example, to obtain a converged value of anomalous Nernst conductivity in PrAlGe \cite{destraz:2020} using {\tt wannier19} (early version of \wBerri), the calculation took approximately 30 hours on the same computation node. 
Estimates predict that the same calculation with \postwx could take several months. 
Thus it is the case where the numerical advance not only saves time, but also brings the calculation from the unreasonably time-consuming area, where most people would avoid working, to a reasonably feasible computation time. 


\section{summary}
In this article I have presented a series of methods that  boost the performance of Wannier interpolation to a new level. The methods are implemented in the  new Python code \wanBerri. 
It is important to note that the mixed Fourier transform and the optimization of MDRS method and Fermi-level iteration, while giving a large computational advantage, do not affect the result within machine precision. Hence the new code can be easily benchmarked with the established \postwx code. 
The code not only allows to perform high-speed and high-precision calculations of AHC and a palette of other properties,  it also serves as a platform for implementating new functionalities involving Wannier interpolation. Thus it has a potential to become a new community code. 
Interestingly, the code uses the same routines to perform calculations both based on WFs and tight-binding models.
Finally, in combination with recent advances in automated construction of WFs, it paves a way to high-throughput calculations of properties of solids that require Wannier interpolation. 

\section*{Acknowledgements}
I thank Ivo Souza and Cheol-Hwan Park for useful discussions and comments helping to improve the manuscript.  I also acknowledge  Xiaoxiong Liu, Miguel \'Angel Jim\'enez Herrera, Patrick M. Lenggenhager, Jae-Mo Lihm, Minsu Ghim for fixing bugs and further development of the code, which will be described elsewhere. 
I acknowledge support from the Swiss National Science Foundation (grant number: PP00P2\_176877), the NCCR MARVEL  and the European Union’s Horizon 2020 research and innovation program (ERC-StG-Neupert-757867-PARATOP). 

\appendix

\section{Functionality implemented in  \wanBerri \label{sec:capabilities}}
This appendix outlines the functionality that is currently implemented in \wanBerri. For more detailed and updated information please refer to the online documentation \cite{wberri-org}

\subsection{Integration}
The code may be used to evaluate the following quantities, represented as Brillouin zone integrals:
\begin{itemize}
\item {\tt 'ahc' : } intrinsic anomalous Hall conductivity $\sigma_{\alpha\beta}^{\rm AHE}$ \cite{Nagaosa-Hall} via \eqref{eq:AHC};
\item Anomalous Nernst conductivity \cite{Xiao-Nernst} $\alpha_{\alpha\beta}^{\rm ANE}$ may be obtained from $\sigma_{\alpha\beta}(\epsilon)^{\rm AHE}$ evaluated over a dense grid of Fermi levels $\epsilon$
\begin{equation}
\alpha_{\alpha\beta}^{\rm ANE}=-\frac{1}{e}\int d\varepsilon \frac{\partial f}{\partial\varepsilon}\sigma_{\alpha\beta}(\varepsilon)\frac{\varepsilon-\mu}{T}, \label{eq:ANE}
\end{equation}
where $f(\varepsilon)=1/\left(1+e^\frac{\varepsilon-\mu}{k_{\rm B}T}\right)$;

\item {\tt 'Morb' : } orbital magnetization
\begin{multline}
    M^\gamma_n(\kk)=\frac{e}{2\hbar}\Im\abc\int\dk\sum_n^{\rm occ}\Bigl[\\
    \me{\partial_a u_{n\kk}}{H_\kk+E_{n\kk}-2E_F}{\partial_b u_{n\kk}}\Bigr];
\end{multline}
\item {\tt 'berry\_dipole' : } berry curvature dipole 
\begin{equation}
D_{\alpha\beta}(\mu)=\int\dk \sum_n^{\rm occ} \partial_\alpha \Omega_n^{\beta}, 
\label{eq:berry-dipole}
\end{equation}
which describes nonlinear Hall effect   \cite{Sodemann-NLHE-prl2015};
\item {\tt 'gyrotropic\_Korb'} and {\tt gyrotropic\_Kspin : } gyrotropic magnetoelectric effect (GME) \cite{zhong-GME-prl2016} tensor (orbital and spin contributions):
\begin{equation}
K_{\alpha\beta}(\mu)=\int\dk\sum_n^{\rm occ}  \partial_\alpha m_n^{\beta} ; \label{eq:gyro-K}
\end{equation}
\item {\tt 'conductivity\_Ohmic' : } ohmic conductivity within the Boltzmann transport theory in constant relaxation time ($\tau$) approximation:
\begin{eqnarray}
\sigma_{\alpha\beta}^{\rm Ohm}(\mu)&=&\tau\int\dk\sum_n^{\rm occ} \partial_\alpha E_{n\kk}\partial_\beta E_{n\kk} \delta(E_{n\kk}-\mu) =\nonumber\\
&=&\tau\int\dk\sum_n^{E_{n\kk}<\mu} \partial^2_{\alpha\beta} E_{n\kk}; \label{eq:ohmic}
\end{eqnarray}
\item {\tt 'dos' : } density of states $n(E)$;
\item {\tt 'cumdos' : } cumulative density of states 
\begin{equation}
    N(E) = \int\limits_{-\infty}^En(\epsilon)d\epsilon.
    \label{eq:cDOS}
\end{equation}
\end{itemize}

\subsection{Tabulating}
\begin{figure}
    \centering
    \includegraphics[width=\linewidth]{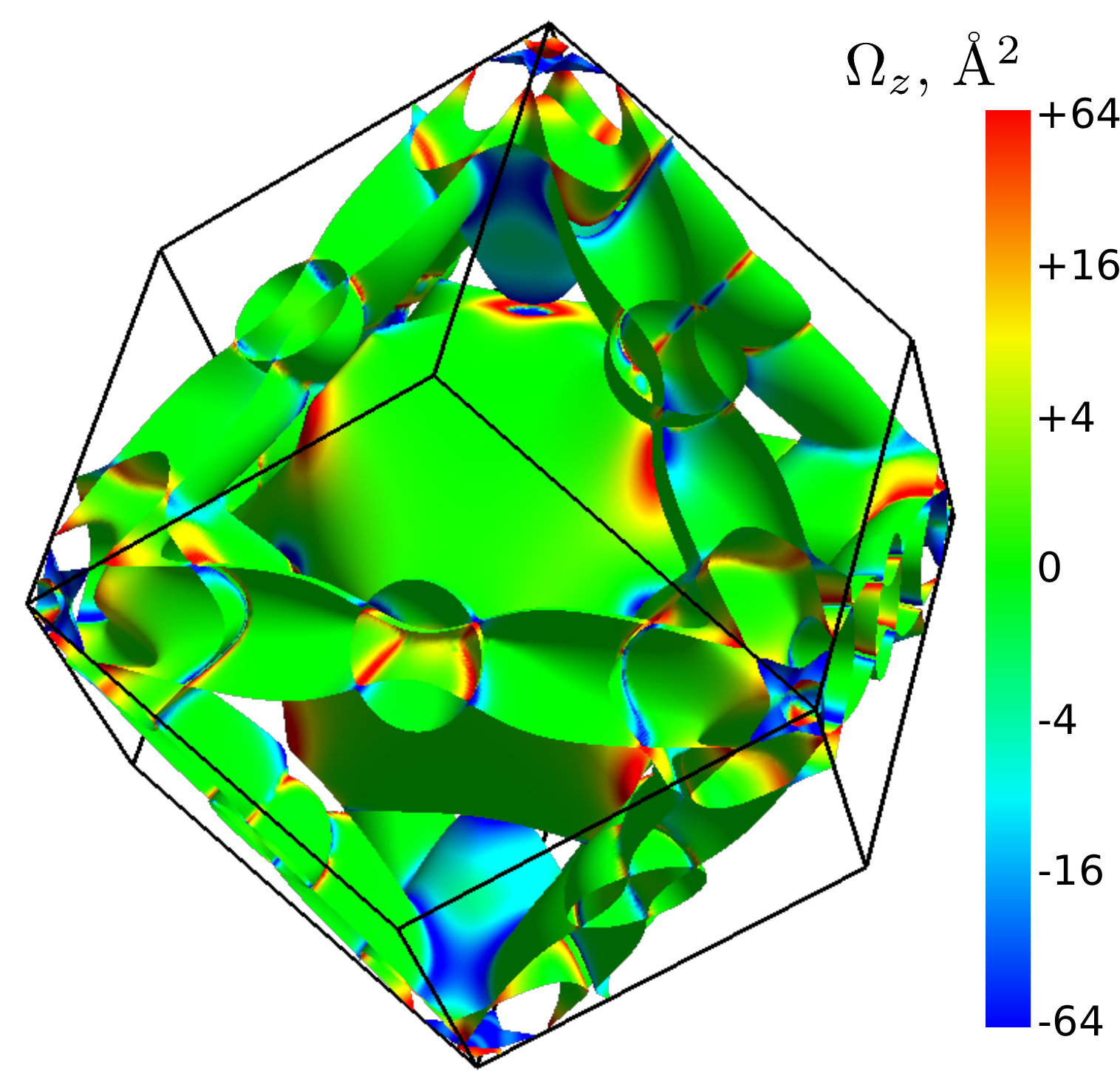}
    \caption{Fermi surface of bcc iron, colored by the Berry curvature $\Omega_z$. Figure produced using {\tt FermiSurfer} \cite{fermisurfer-paper}. }
    \label{fig:Fe-frmsf}
\end{figure}

\wBerri can also tabulate certain band-resolved quantities over the Brillouin zone. This feature is called by, e.g. 
\begin{lstlisting}[language=Python,name={tutorial}]
WB.tabulate(system, grid,
             quantities=["berry"],
             fout_name="Fe",
             numproc=16,
             ibands=np.arange(4,10),
             Ef0=12.610)
\end{lstlisting}
which will produce files {\tt Fe\_berry-?.frmsf}, containing the Energies and Berry curvature of bands {\tt 4-9}  (band counting starts from zero). 
The format of the files allows to be directly passed to the {\tt FermiSurfer} visualization tool \cite{fermisurfer-paper} which can produce a plot like Fig~\ref{fig:Fe-frmsf}.  Transformation of files to other visualization software is straightforward.

Currently the following quantities are available to tabulate:
\begin{itemize}
\item {\tt 'berry' : } Berry curvature
\begin{equation}
\Omega^\gamma_n(\kk)=-\abc\Im\ip{\partial_\alpha u_{n\kk}}{\partial_\beta u_{n\kk}};
    \end{equation}
\item {\tt 'morb' : }  orbital moment of Bloch states
\begin{equation}
    m^\gamma_n(\kk)=\frac{e}{2\hbar}\abc\Im\me{\partial_\alpha u_{n\kk}}{H_\kk-E_{n\kk}}{\partial_\beta u_{n\kk}};
\end{equation}
\item {\tt 'spin' : } the expectation value of the Pauli operator
\begin{equation}
    \mathbf{s}_n(\kk)=\me{u_{n\kk}}{\hat{\bf \sigma}}{u_{n\kk}};
\end{equation}
\item {\tt 'V' : } the band gradients $\nabla_\kk E_{n\kk}$.
\end{itemize}

\section{Evaluation of additional matrix elements \label{sec:matel}}

\begin{figure}[tp!]
    \centering
    \includegraphics[width=\columnwidth]{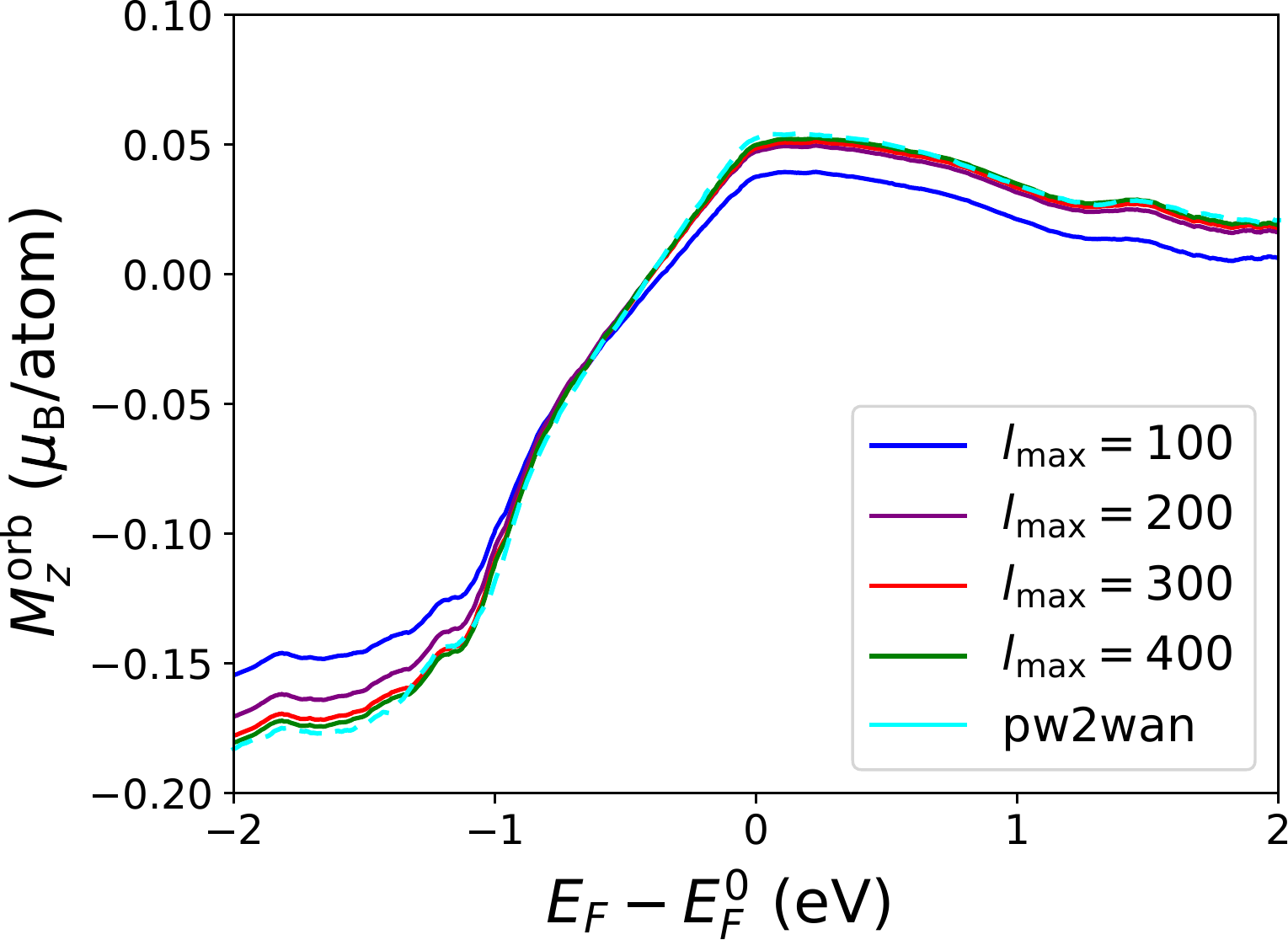}
    \caption{Orbital magntization of bcc Fe as a function of the Fermi level $E_F$ reative to the pristine Fermi level $E_F^0$, evaluated using the \texttt{.uHu} file computed with \texttt{pw2wannier90.x} (dashed line) and using the \texttt{wannierberri.mmn2uHu} interface (solid lines) with different summation limit $l_{\rm max}$ in \eqref{eq:Cmnq-sum}. }
    \label{fig:mmn2uHu}
\end{figure}

Wannier interpolation starts from certain matrix elements defined on the \textit{ab initio} ($\qq$) grid. 
Those matrix elements should be evaluated within the \textit{ab initio} code, namely within its interface to Wannier90. 
However, only QuantumEspresso \cite{QE-2020} has the most complete interface {\tt pw2wannier90.x}. 
The other codes provide only the basic interface, which includes the eigenenergies $E_{n\qq}$ ({\tt .eig} file) and overlaps  
\begin{equation}
    M_{mn}^{\mathbf{b}}(\qq)=\ip{u_{m\qq}}{u_{n\qq+\mathbf{b}}}
\end{equation}
(file {\tt .mmn}), where $\mathbf{b}$ vector connects neighbouring $\qq$-points. 
This information allows to interpolate the band energies (and their derivatives of any order) as well as Berry connections  \cite{yates-2007} and Berry curvature  \cite{wang-prb06}. However, to evaluate the orbital moment of a Bloch state, one needs matrix elements of the Hamiltonian  \cite{lopez-prb12} ({\tt .uHu} file)
\begin{equation}
    C_{mn}^{\mathbf{b}_1,\mathbf{b}_2}(\qq)=\me{u_{m\qq+\mathbf{b}_1}}{\hat{H}_\qq}{u_{n\qq+\mathbf{b}_2}}.
    \label{eq:Cmnq}
\end{equation}
The evaluation of \eqref{eq:Cmnq} is very specific to the details of the \textit{ab initio} code, and implemented only in {\tt pw2wannier90.x} and only for norm-conserving pseudopotentials.  To enable the study of properties related to the orbital moment with other \textit{ab initio} codes, the following workaround may be employed. By inserting a complete set of Bloch states at a particular $\qq$ point $\mathbbm{1}=\sum_l^\infty \ket{u_{l\qq}}\bra{u_{l\qq}}$ we can rewrite \eqref{eq:Cmnq}  as 
\begin{equation}
    C_{mn}^{\mathbf{b}_1,\mathbf{b}_2}(\qq)\approx\sum_l^{l_{\rm max}}  \left(M_{lm}^{\mathbf{b}_1}(\qq)\right)^* E_{l\qq}   M_{ln}^{\mathbf{b}_2}(\qq).
    \label{eq:Cmnq-sum}
\end{equation}
This equation is implemented within the {\tt wannierberri.mmn2uHu} submodule, which allows to generate the {\tt .uHu} file out of {\tt .mmn} and  {\tt .eig} files.
The equality in \eqref{eq:Cmnq-sum} is exact only in the limit $l_{\rm max}\to\infty$ and infinitely large basis set for the wavefunctions representation. 
So in practice one has to check convergence for a particular system. As an example the bandstructure of bcc Fe was calculated based on the QE code and a norm-sonserving pseudopotential from the PseudoDojo library\cite{pseudodojo,pseudodojo-normconserving}. Next, the orbital magnetization was calculated using the \texttt{.uHu} file computed with \texttt{pw2wannier90.x} and using the \texttt{wannierberri.mmn2uHu} interface with different summation limit $l_{\rm max}$ in \eqref{eq:Cmnq-sum}. As can be seen in  Fig.~\ref{fig:mmn2uHu} already $l_{\rm max}=200$ (corresponding energy $\sim 230$ eV) yields a result very close to that of \texttt{pw2wannier90.x}. However one should bear in mind that convergence depends on many factors, such as as choice of WFs and pseudopotentials. In particular, for tellurium we observed \cite{tsirkin-prb18} that including only a few bands above the highest $p$-band is enough to obtain accurate results.  However for iron, using a pseudopotential shipped with the examples of Wannier90, we failed to reach convergence even with $l_{\rm max}=600$. 


To interpolate the spin operator expectation value, the  matrix $s_{mn}(\qq)=\me{u_{m\qq}}{\hat{\sigma}}{u_{n\qq}}$ is needed. To facilitate study of spin-dependent properties within VASP \cite{vasp} code, a submodule {\tt wannierberri.vaspspn} is included, which computes $S_{mn\qq}$ based on the normalized pseudo-wavefunction read from the {\tt WAVECAR} file. Note that the use of pseudo-wavefunction instead of the full PAW \cite{PAW} wavefunction is an approximation, which however in practice gives a rather accurate interpolation of spin. 

The {\tt mmn2uHu} and {\tt vaspspn} modules were initially developed and used  in  \cite{tsirkin-prb18} as separate scripts, but were not published so far. Now they are included in the \wanBerri package with a hope of being useful for the community.

\section{Software and data availability}
All software used and developed in this article is open-source and available for free. \wanBerri \cite{wberri-org} is available via {\tt pip} \cite{pypi-wberri} and GitHub \cite{github-wberri}. 
External libraries used in \wanBerri include {\tt NumPy} \cite{numpy}, {\tt SciPy} \cite{scipy}, {\tt pyFFTW} \cite{pyFFTW} wrapper of FFTW3  library \cite{FFTW05} and  {\tt lazy-propery} \cite{lazy-property}.
Wannier90 \cite{MOSTOFI2008685,MOSTOFI20142309,Pizzi_2020} and QuantumEspresso \cite{QE-2020} are available at \cite{wannier.org,github-w90} and \cite{qe.org}. 
Figures were produced using {\tt matplotlib} \cite{matplotlib} (Figs.~\ref{fig:refinement}-\ref{fig:timing-fscan} and \ref{fig:mmn2uHu}) and FermiSurfer\cite{fermisurfer-paper} (Fig.~\ref{fig:Fe-frmsf}), and also with help of Inkscape \cite{Inkscape}. 

All input files needed to generate the WFs to reproduce the example of the manuscript are part of tutorial\#18 of Wannier90, and are available at  \cite{wannier.org,github-w90}. 


\bibliography{bib}

\end{document}